\documentclass[11pt]{iopart}

\expandafter\let\csname equation*\endcsname\relax 
\expandafter\let\csname endequation*\endcsname\relax 
\usepackage{amsmath,amssymb}
\usepackage{fullpage,multirow,times}
\usepackage{graphicx}
\usepackage{color}
\usepackage{empheq} 
\usepackage{listings}
\usepackage{array, tabularx}
\usepackage{makecell}
\usepackage{geometry}

\usepackage{soul} 

\usepackage{mathtools} 

\usepackage{mathrsfs}
\usepackage[mathscr]{euscript}

\usepackage{verbatim}
\usepackage{enumitem} 

\usepackage{etex}
\usepackage{cases}

\usepackage{scalerel}

\setlist{nolistsep}
\usepackage{float}

\usepackage{bigstrut}
\usepackage[english]{babel} 
\usepackage{booktabs} 
\usepackage[font=small]{caption}
\usepackage{setspace}

\usepackage{natbib}
\usepackage{hyperref}
\setcitestyle{authoryear,open={(},close={)}}

\usepackage{pdfpages}

\usepackage{lipsum} 
\usepackage[svgnames]{xcolor} 

\usepackage{accents}

\DeclareMathAlphabet{\mathpzc}{OT1}{pzc}{m}{it}

\definecolor{wisconsin-red}{rgb}{0.6,0,0}
\definecolor{darkgreen}{rgb}{0.2,0.6,0.2}
\definecolor{maroon}{rgb}{0.5, 0.0, 0.0}
\definecolor{violet}{rgb}{0.75, 0.0, 1.0}
\definecolor{lightgray}{gray}{0.9}
\definecolor{navyblue}{rgb}{0.0, 0.0, 0.35}
\definecolor{darkmidnightblue}{rgb}{0.0, 0.2, 0.4}
\definecolor{Gray}{gray}{0.75}
\definecolor{darkgreen}{rgb}{0,0.5,0}

\newcommand{\disps}{\displaystyle}

\newcommand*\boxSizeOfMax[1]{\makebox[\widthof{max}][c]{#1}}
\newcommand{\sthat}{\boxSizeOfMax{s.t.}}

\newcommand{\hhsp}{\-\hspace}
\newcommand{\hhspe}{\-\hspace{0.8cm}}

\newcommand{\vvspnt}{\vspace{-0.3cm}}

\newcommand{\tcolK}{\textcolor{black}}

\newcommand{\bit}{\begin{itemize}}
\newcommand{\eit}{\end{itemize}}
\newcommand{\benn}{\begin{enumerate}}
\newcommand{\eenn}{\end{enumerate}}
\newcommand{\bfg}{\begin{figure}[H]}
\newcommand{\efg}{\end{figure}}
\newcommand{\bcn}{\begin{center}}
\newcommand{\ecn}{\end{center}}

\newcommand{\ignore}[1]{}

\renewcommand{\arraystretch}{1.3}

\setlist[itemize,2]{label=$\textcolor{navyblue}{\circ}$} 
\setlist[itemize,4]{label=$\textcolor{navyblue}{\checkmark}$}
\setlist[itemize,3]{label=$\textcolor{navyblue}{\mathsmaller{\diamondsuit}}$}
\setlist[itemize,1]{label=$\textcolor{navyblue}{\mathsmaller{\mathsmaller{\square}}}$}

\usepackage{multirow}
\usepackage{colortbl}

\usepackage{float}
\usepackage[export]{adjustbox}
\usepackage[caption = false]{subfig}

\usepackage[norelsize, boxed]{algorithm2e}
\usepackage{algorithmic}
\usepackage{relsize} 

\usepackage{lscape}
\usepackage{pdflscape}
\usepackage{afterpage}
\usepackage[flushleft]{threeparttable}

\usepackage{tikz}
\usepackage{tikz-qtree}
\usetikzlibrary{plotmarks}
\usetikzlibrary{shapes,shadows,arrows,snakes,fit}
\usetikzlibrary{decorations.pathmorphing}
\usetikzlibrary{decorations.markings}
\usetikzlibrary{calc} 
\usetikzlibrary{positioning,chains,scopes}

\usetikzlibrary{shapes.geometric,shapes.arrows,decorations.pathmorphing}
\usetikzlibrary{matrix,chains,scopes,positioning,arrows,fit}

\usepackage[toc]{appendix}
\usepackage{xpatch}
\let\appendixpagenameorig\appendixpagename
\renewcommand{\appendixpagename}{\sffamily\appendixpagenameorig}

\newcommand{\isoVar}{z}
\newcommand{\bigM}{M}
\newcommand{\UBNumIsocenter}{N}

\newcommand{\structure}{\ell}
\newcommand{\setStructure}{\mathcal{L}}

\newcommand{\beamWeight}{w}

\newcommand{\doseMat}{\Delta}

\newcommand{\dose}{d}

\newcommand{\doseOver}{o}

\newcommand{\doseUnder}{u}
\newcommand{\setVoxel}{\mathcal{V}}
\newcommand{\voxel}{v}
\newcommand{\setIsocenter}{\mathcal{I}}

\newcommand{\isocenter}{\theta}
\newcommand{\setSector}{\mathcal{S}}
\newcommand{\sector}{s}
\newcommand{\setSectorSize}{\mathcal{K}}
\newcommand{\sectorSize}{k}
\newcommand{\setTumor}{\mathcal{T}}
\newcommand{\tumor}{t}
\newcommand{\setOAR}{\mathcal{C}}
\newcommand{\oar}{c}
\newcommand{\targetDose}{\pi}

\newcommand{\objWeightLP}{\alpha}
\newcommand{\objWeightLPTumUnder}{\underline{\objWeightLP}}
\newcommand{\objWeightLPTumOver}{\bar{\objWeightLP}}

\newcommand{\sdoTime}{\mathcal{Q}}
\newcommand{\coefBOT}{\varrho}

\newcommand{\botOpt}{q}

\newcommand{\dualMaxVar}{\lambda}
\newcommand{\dualMinVar}{\gamma}

\newcommand{\dualBotOpt}{\nu}

\newcommand{\masterVar}{\eta}
\newcommand{\isoCoef}{\beta}

\newcommand{\modelSDIOone}{\text{[$\mathrm{SDIO}$]}}

\newcommand{\modelLPone}{\text{[$\mathrm{LP}$]}}

\newcommand{\modelDualLPone}{\text{[$\mathrm{DualLP}$]}}

\newcommand{\modelSDIOMPone}{\text{[$\mathrm{MP}$]}}

\DeclarePairedDelimiterX{\norm}[1]{\lVert}{\rVert}{#1}

\setitemize{itemsep=5pt}

\makeindex
\date{}

\begin{document}

\title{Simultaneous optimization of isocenter locations and sector duration in radiosurgery}

\author{M. Cevik$^1$, D. Aleman$^2$, Y. Lee$^1$, A. Berdyshev$^2$, H. Nordstrom$^3$, S. Riad$^3$, A. Sahgal$^1$, M. Ruschin$^{1,2}$}

\address{$^1$ Department of Radiation Oncology, Sunnybrook Health Sciences Center, University of Toronto, ON, Canada\\
$^2$ Department of Mechanical and Industrial Engineering, University of Toronto, ON, Canada\\
$^3$ Elekta Instrument, AB, Sweden}
\ead{mark.ruschin@sunnybrook.ca}


\begin{abstract}
Stereotactic radiosurgery is an effective technique to treat brain tumors for which several inverse planning methods may be appropriate. We propose an integer programming model to simultaneous sector duration and isocenter optimization (SDIO) problem for Leksell Gamma Knife{\textregistered} Icon{\texttrademark} (Elekta, Stockholm, Sweden) to tractably incorporate treatment time. We devise a Benders decomposition scheme to solve the SDIO problem to optimality. The performances of our approaches are assessed using anonymized data from eight previously treated cases, and obtained treatment plans are compared against each other and against the clinical plans. The plans generated by our SDIO model all meet or exceed clinical guidelines while demonstrating high conformity.


\end{abstract}

\newpage
\section{Introduction}
Stereotactic radiosurgery (SRS) is an effective method for treatment of brain tumors, vascular malformations, and other conditions such as tremors. During SRS, a large amount of radiation is delivered to tumors through the intact skull without damaging the surrounding normal brain tissue. SRS aims to obtain highly conformal dose distributions so that dose is delivered to target structures with high precision while sparing the surrounding healthy tissues, thus increasing the patient's survival chance and improving the quality of life after the treatment.

One advanced SRS system is Leksell Gamma Knife{\textregistered} (LGK) (Elekta, Stockholm, Sweden) Icon{\texttrademark}. Icon{\texttrademark} has a single collimator with eight detached sectors that can either be robotically driven to three different collimator sizes (4, 8, and 16 mm) or be blocked. During SRS, the patient lies on a couch and radiation is emitted by 192 cobalt-60 sources to the targets where the intersection point of all the beams is called the \textit{isocenter}. The radiation delivery follows a step-and-shoot manner, i.e., the couch is stationary during delivery of radiation at each isocenter location, and only moves when the beam is blocked. To fully leverage the robotic capability of the delivery system, an automated approach for planning is required. The clinical goals for such an automated approach are: (1) satisfy or exceed the dosimetric clinical objectives (e.g., coverage, conformality, healthy tissue dose); (2) achieve (1) with minimal irradiation time (i.e., beam-on time); and (3) achieve (1) and (2) with a plan as efficient as possible, including minimizing the number of isocenters and shots required.

Most of the current literature has separated the task of isocenter placement from the task of sector duration optimization (SDO). In SDO, as in fluence map optimization in Intensity Modulated Radiation Therapy, the aim is to find the intensities (or irradiation times) of each beamlet in a fixed set of beams. Previous studies on SDO mainly focus on variants of convex quadratic penalty models \citep{Oskoorouchi2011, Ghobadi2012, Ghaffari2012}. In order to determine the isocenter locations, \citet{Ghobadi2012} propose an algorithm that combines the well-known grassfire algorithm \citep{Wagner2000} with a sphere packing routine, which iteratively uses the deepest voxels in the tumor as candidates and each candidate isocenter is then assigned to a sphere of voxels that it covers. In another study, \citet{Doudareva2015} introduce an isocenter selection algorithm based on the skeletonization techniques, which identifies clusters of each tumor's skeleton using distance coding methods. \citet{Ghaffari2017} propose an integer programming model to combine the isocenter selection and SDO phases of the treatment planning. All of the aforementioned studies report high treatment plan quality in terms of satisfying the dosimetric clinical objectives with limited attention paid to satisfying clinical goals (2) and (3). In a recent study, \citet{Cevik2018} investigate three modeling approaches for the SDO: linear programming, convex quadratic penalty approach, and convex moment-based approach. Although all three approaches yield plans satisfying the dosimetric clinical objectives, their findings suggest that the linear programming tends to yield the lowest beam-on time (BOT).

In this paper, we consider the simultaneous optimization of sector duration and isocenter locations (SDIO) in LGK Icon{\texttrademark} with the goal of presenting a fully automated approach that satisfies the clinical goals stated above. Specifically, we start with a linear programming formulation for SDO that aims to find conformal treatment plans with a low BOT. Then, we formulate an integer programming model for SDIO and propose a Benders decomposition scheme to improve the solvability of the model.

\section{Methods}\label{sec:methods}
We assume that a candidate set of isocenter positions are already determined (e.g., using a grassfire and sphere packing algorithm \citep{Ghobadi2012} or random sampling) and use these isocenters to determine the irradiation time from each sector at each isocenter location. In our models, we aim to satisfy the dose requirements for targets, organs-at-risk (OARs), and additional avoidance structures such as healthy tissues surrounding the targets (i.e., rings). For the sake of simplicity, we consider these rings to be OARs as well, however, in reality an OAR is usually a specific organ such as brainstem or cochlea, which may have strict dose restrictions.

Table \ref{tb:not} summarizes the notation used throughout the paper. Each structure $\structure \in \setStructure = \{\setTumor, \ \setOAR\}$, where $\setTumor$ and $\setOAR$ denote the set of tumors and OARs, respectively, consists of a set of voxels $\setVoxel_{\structure}$. We use $\setIsocenter, \ \setSector$ and $\setSectorSize$ to represent the set of isocenters, sectors and collimator sizes, respectively. Let $\doseMat_{\structure \voxel\isocenter\sector\sectorSize}$ represent delivered dose per unit time to voxel $\voxel\in \setVoxel_{\structure}$ of structure $\structure \in \setStructure$ from isocenter $\isocenter \in \setIsocenter$ and sector $\sector \in \setSector$ at collimator size $\sectorSize \in \setSectorSize$, and $\beamWeight_{\isocenter \sector \sectorSize}$ represent \textit{time duration of radiation delivery} to isocenter location $\isocenter \in \setIsocenter$ from sector $\sector\in \setSector$ at collimator size $\sectorSize \in \setSectorSize$. Then, we can obtain the dose delivered to voxel $\voxel \in \setVoxel_{\structure}$ in structure $\structure \in \setStructure$ as follows:
	\begin{equation}
	\label{e:mc:vox1} \dose_{\structure \voxel} = \sum_{\isocenter \in \setIsocenter} \sum_{\sector \in \setSector} \sum_{\sectorSize \in \setSectorSize} \ \doseMat_{\structure \voxel\isocenter\sector\sectorSize} \beamWeight_{\isocenter \sector \sectorSize}.
	\end{equation}
	
\setlength{\tabcolsep}{6pt}
\begin{table}[h]
\centering
\caption{Model notation}
\scalebox{0.95}{
\begin{tabular}{lll}
\midrule
Notation &  & Description \\
\midrule
$\isocenter \in \setIsocenter, \ \sector \in \setSector, \ \sectorSize \in \setSectorSize$ & & Isocenter, sector, and sector size, and their corresponding sets\\
$\structure \in \setStructure, \ \tumor \in \setTumor,\ \oar \in \setOAR$ & & Structure, tumor, and OAR, and their corresponding sets\\
$\beamWeight_{\isocenter\sector\sectorSize}$ & & \makecell[l]{Irradiation time from isocenter $\isocenter$ sector $\sector$ and sector size $\sectorSize$}\\
$\voxel \in \setVoxel_{\structure}$ & & Voxel and set of voxels in structure $\structure$\\

$\doseMat_{\structure \voxel\isocenter\sector\sectorSize}$ & & \makecell[l]{Dose received by voxel $\voxel \in \setVoxel_\structure$ from isocenter $\isocenter$ sector $\sector$ and sector size $\sectorSize$}\\
$\dose_{\structure \voxel}$ & & Dose received by voxel $\voxel$ in structure $\structure$\\
$\targetDose_{\structure}, \ \targetDose_{\structure}^{\max}$ & & Target dose and maximum allowed dose for structure $\structure$\\[0.5em]

\multicolumn{3}{c}{\textbf{LP/IP model components}}\\
\cmidrule(lr){1-3}
$\doseOver_{\structure \voxel}, \ \doseUnder_{\structure \voxel}$ & & \makecell[l]{LP model variables representing overdose and underdose amounts for\\ voxel $\voxel$ in structure $\structure$ for the LP model} \\
$\botOpt_{\isocenter}$ & & \makecell[l]{LP model variable representing the BOT value for isocenter $\isocenter$}\\
$\dualMaxVar_{\structure \voxel}, \ \dualMinVar_{\structure \voxel}$ & & \makecell[l]{DualLP model variables corresponding to overdose and underdose\\ constraints in LP}\\
$\dualBotOpt_{\isocenter\sector}$ & & \makecell[l]{DualLP model variables corresponding to BOT constraints in LP}\\
$\isoVar_{\isocenter}$ & & \makecell[l]{SDIO model variables representing selection of isocenter $\isocenter$}\\
$\masterVar$ & & \makecell[l]{SDIO MP model variables for approximating the subproblem objective weights}\\
$\objWeightLP_{\structure}, \ \objWeightLPTumUnder_{\structure},\ \objWeightLPTumOver_{\structure}$ & & \makecell[l]{LP model objective weights: weight assigned for delivered dose,\\ underdose and overdose for structure $\structure$}\\
$\UBNumIsocenter$ & & Upper bound on the number of isocenters used\\
$\isoCoef$ & & Time spent switching between two isocenters\\
$\coefBOT$ & & BOT component weight parameter\\[0.5em]

\midrule
\end{tabular}
}
\label{tb:not}
\end{table}

SRS protocols usually recommend that at least 98\% of the tumor volume should receive the prescribed dose $\targetDose_{\tumor}, \ \tumor \in \setTumor$. Moreover, the maximum delivered dose to the tumors in SRS, $\targetDose_{\tumor}^{\max}, \ \tumor \in \setTumor$, is usually recommended not to exceed two times the prescribed dose, and dose to the healthy organs is recommended to be less than the maximum allowed dose $\targetDose_{\oar}^{\max}, \ \oar \in \setOAR$.

One of the objectives of the sector duration optimization is to minimize the BOT of the treatment plans. Specifically, BOT for an isocenter $\isocenter \in \setIsocenter$ is $\sdoTime_{\isocenter} = \max_{\sector \in \setSector} \Big\{ \displaystyle\sum\limits_{\sectorSize \in \setSectorSize} \beamWeight_{\isocenter \sector \sectorSize}\Big\}$. Then, the BOT of the overall treatment plan can be calculated as $\sum_{\isocenter \in \setIsocenter} \sdoTime_{\isocenter}$.

\subsection{LP model for SDO}
We use a linear programming (LP) model for the SDO problem which aims to balance the conflicting goals of the treatment plan such as delivering required dose to targets and minimizing dose to healthy structures around the targets. Due to the complicated nature of radiosurgery, it is not always possible to deliver the required dose to each voxel of the targets without violating the dose limits for the critical structures. Accordingly, we relax the dose requirement constraints in the model by introducing nonnegative underdose and overdose auxiliary variables, namely, $\doseUnder_{\tumor \voxel}$ for underdosing tumor voxel $\voxel \in \setVoxel_{\tumor}, \ \tumor \in \setTumor$, and $\doseOver_{\structure \voxel}$ for overdosing voxel $\voxel \in \setVoxel_{\structure}, \structure \in \setStructure$. We formulate an LP model for SDO as follows:\\
\begin{small}
\begin{subequations}
	\label{e:prlp}
	\begin{align}
	\label{e:prlp0} \hhsp{-0.1cm} \text{\modelLPone}: \min & \ \sum_{\oar\in \setOAR} \objWeightLP_{\oar} \sum_{\voxel \in \setVoxel_{\oar}} (\dose_{\oar \voxel} + \doseOver_{\oar \voxel}) + \sum_{\tumor\in \setTumor} \objWeightLPTumUnder_{\tumor} \sum_{\voxel \in \setVoxel_{\tumor}} \doseUnder_{\tumor \voxel} + \sum_{\tumor\in \setTumor} \objWeightLPTumOver_{\tumor} \sum_{\voxel \in \setVoxel_{\tumor}} \doseOver_{\tumor \voxel} + \tcolK{\coefBOT \sum_{\isocenter \in \setIsocenter} \botOpt_{\isocenter}} \hhsp{-4.8cm}\\
	\label{e:prlp1} \sthat & \ \dose_{\structure \voxel} = \sum_{\isocenter \in \setIsocenter}\sum_{\sector \in \setSector}\sum_{\sectorSize \in \setSectorSize}\doseMat_{\structure\voxel\isocenter\sector\sectorSize} \beamWeight_{\isocenter\sector\sectorSize} && \quad \voxel \in \setVoxel_{\structure}, \ \structure \in \setStructure\\
	\label{e:prlp2} & \ \dose_{\tumor \voxel} + \doseUnder_{\tumor \voxel}\geq \targetDose_{\tumor} && \quad \voxel \in \setVoxel_{\tumor}, \ \tumor\in \setTumor\\
	\label{e:prlp3} & \ \dose_{\structure \voxel} - \doseOver_{\structure \voxel}\leq \targetDose_{\structure}^{\max} && \quad \voxel \in \setVoxel_{\structure}, \ \structure\in \setStructure\\
	\label{e:prlp4} & \ \tcolK{\botOpt_{\isocenter} \geq \sum_{\sectorSize \in \setSectorSize} \beamWeight_{\isocenter \sector \sectorSize}}, && \quad \tcolK{\isocenter \in \setIsocenter, \ \sector \in \setSector}\\
	\label{e:prlp5} & \ \tcolK{\botOpt_{\isocenter} \geq 0}, \ \beamWeight_{\isocenter\sector\sectorSize} \geq 0  && \quad \isocenter \in \setIsocenter, \ \sector \in \setSector, \ \sectorSize \in \setSectorSize\\
	\label{e:prlp6} & \ \doseUnder_{\structure \voxel} \geq 0, \ \ \doseOver_{\structure \voxel} \geq 0 && \quad \voxel \in \setVoxel_{\structure}, \ \structure\in \setStructure
	\end{align}
	\end{subequations}
\end{small}

\noindent The objective \eqref{e:prlp0} minimizes the weighted average of the underdose and overdose to the structures, dose delivered to the OARs and BOT of the resulting plan, where $\objWeightLP_{\oar}$ represents the weight associated with OAR $\oar \in \setOAR$, while $\objWeightLPTumUnder_{\tumor}$ and $\objWeightLPTumOver_{\tumor}$ represent the weights for the underdose and overdose amounts to tumor $\tumor \in \setTumor$, respectively. Constraints \eqref{e:prlp1} determine the dose delivered to each voxel at each structure. Constraints \eqref{e:prlp2} and \eqref{e:prlp3} relate the underdose and overdose variables, respectively, with the delivered dose amounts and given dose limits. Constraints \eqref{e:prlp4} along with $\coefBOT \sum_{\isocenter \in \setIsocenter} \botOpt_{\isocenter}$ terms in the objective function are used to control the BOT of the resulting plan noting that $\botOpt_{\isocenter}, \ \isocenter \in \setIsocenter$, variables correspond to BOT for each isocenter. Note that the weight parameters, $\{\objWeightLP_{\oar}, \ \objWeightLPTumUnder_{\tumor}, \ \objWeightLPTumOver_{\tumor}\}$, are not known in advance, and thus, may be adjusted to balance the conflicting goals of the SDO such as minimizing the BOT and ensuring that tumors receive the prescribed dose.

The dual of $\modelLPone$ can be formulated as follows:
\begin{small}
\begin{subequations}
		\begin{align}
	\label{e:dlpr0} \hhsp{-0.4cm} \text{\modelDualLPone}: \max & \ -\sum_{\structure \in \setStructure} \sum_{\voxel \in \setVoxel_{\structure}} \targetDose_{\structure}^{\max} \dualMaxVar_{\structure \voxel} + \sum_{\tumor \in \setTumor} \sum_{\voxel \in \setVoxel_{\tumor}} \targetDose_{\tumor} \dualMinVar_{\tumor \voxel} \hhsp{-4.8cm}\\
		\label{e:dlpr1} \sthat & \ \sum_{\tumor \in \setTumor} \sum_{\voxel \in \setVoxel_{\tumor}} \doseMat_{\tumor\voxel\isocenter\sector\sectorSize} \dualMinVar_{\tumor \voxel} - \sum_{\structure \in \setStructure} \sum_{\voxel \in \setVoxel_{\structure}} \doseMat_{\structure\voxel\isocenter\sector\sectorSize} \dualMaxVar_{\structure \voxel} \tcolK{- \dualBotOpt_{\isocenter\sector}} \leq \sum_{\oar \in \setOAR} \sum_{\voxel \in \setVoxel_{\oar}} \objWeightLP_{\oar} \doseMat_{\oar\voxel\isocenter\sector\sectorSize} && \isocenter \in \setIsocenter, \ \sector \in \setSector, \ \sectorSize \in \setSectorSize\\
	\label{e:dlpr6} & \ \tcolK{\sum_{\sector\in \setSector} \dualBotOpt_{\isocenter\sector} \leq \coefBOT} && \tcolK{\isocenter\in \setIsocenter}\\
	\label{e:dlpr2} & \ \dualMaxVar_{\oar \voxel} \leq \objWeightLP_{\oar} && \oar \in \setOAR, \ \voxel \in \setVoxel_{\oar}\\
	\label{e:dlpr3} & \ \dualMaxVar_{\tumor \voxel} \leq \objWeightLPTumOver_{\tumor} && \tumor \in \setTumor, \ \voxel \in \setVoxel_{\tumor}\\
	\label{e:dlpr4} & \ \dualMinVar_{\tumor \voxel} \leq \objWeightLPTumUnder_{\tumor} && \tumor \in \setTumor, \ \voxel \in \setVoxel_{\tumor}\\
	\label{e:dlpr5} & \ \dualMinVar_{\structure \voxel} \geq 0, \ \ \dualMaxVar_{\structure \voxel} \geq 0 && \voxel \in \setVoxel_{\structure}, \ \structure\in \setStructure\\
	\label{e:dlpr7} & \ \tcolK{\dualBotOpt_{\isocenter\sector} \geq 0} && \tcolK{\isocenter\in \setIsocenter,\ \sector \in \setSector}
	\end{align}
	\end{subequations}
	\end{small}

\noindent Note that $\modelLPone$ has $\Big(\sum_{\structure \in \setStructure} |\setVoxel_{\structure}| + \sum_{\tumor \in \setTumor} |\setVoxel_{\tumor}| + |\setIsocenter|\times|\setSector| \Big)$ many constraints (excluding the variable bounds) and $\Big(\sum_{\structure \in \setStructure} |\setVoxel_{\structure}| + \sum_{\tumor \in \setTumor} |\setVoxel_{\tumor}| + |\setIsocenter|\times|\setSector|\times|\setSectorSize| + |\setIsocenter|\Big)$ many variables. On the other hand, observing that constraints \eqref{e:dlpr2}, \eqref{e:dlpr3}, and \eqref{e:dlpr4} can be taken as variable bounds, $\modelDualLPone$ has $\Big(|\setIsocenter|\times|\setSector|\times|\setSectorSize| + |\setIsocenter|\Big)$ many constraints and $\Big(\sum_{\structure \in \setStructure} |\setVoxel_{\structure}| + \sum_{\tumor \in \setTumor} |\setVoxel_{\tumor}| + |\setIsocenter|\times|\setSector|\Big)$ many variables. As such, depending on the isocenter and voxel counts, $\modelDualLPone$ might have significantly less constraints and variables. Therefore, we recognize that $\modelDualLPone$ might be significantly easier to solve compared to $\modelLPone$.

\subsection{IP model for SDIO}
We formulate an integer programming model (IP) to simultaneously determine the isocenter locations and the irradiation times from each isocenter and sector for each collimator size. In the IP model, we assume that the set of \textit{eligible} isocenter locations, $\setIsocenter$, is predetermined and we enforce a limit on the number of isocenters used in the treatment plan. Let variable $\isoVar_{\isocenter} \in \{0,1\}$ indicate whether or not isocenter $\isocenter \in \setIsocenter$ is selected to be used in the treatment. Then, we formulate the IP model for SDIO as follows:
\begin{small}
\begin{subequations}
\label{e:sdior}
\begin{align}
	\label{e:sdior0} \hhsp{-0.2cm} \text{\modelSDIOone}: \min & \ \sum_{\oar\in \setOAR} \objWeightLP_{\oar} \sum_{\voxel \in \setVoxel_{\oar}} (\dose_{\oar \voxel} + \doseOver_{\oar \voxel}) + \sum_{\tumor\in \setTumor} \objWeightLPTumUnder_{\tumor} \sum_{\voxel \in \setVoxel_{\tumor}} \doseUnder_{\tumor \voxel} + \sum_{\tumor\in \setTumor} \objWeightLPTumOver_{\tumor} \sum_{\voxel \in \setVoxel_{\tumor}} \doseOver_{\tumor \voxel} + \tcolK{\coefBOT \sum_{\isocenter \in \setIsocenter} \big(\botOpt_{\isocenter} + \isoCoef \isoVar_{\isocenter}\big)} \hhsp{-4.8cm}\\
	\label{e:sdior1} \sthat & \ \dose_{\structure \voxel} = \sum_{\isocenter \in \setIsocenter}\sum_{\sector \in \setSector}\sum_{\sectorSize \in \setSectorSize}\doseMat_{\structure\voxel\isocenter\sector\sectorSize} \beamWeight_{\isocenter\sector\sectorSize} && \quad \voxel \in \setVoxel_{\structure}, \ \structure \in \setStructure\\
	\label{e:sdior2} & \ \dose_{\structure \voxel} - \doseOver_{\structure \voxel}\leq \targetDose_{\structure}^{\max} && \quad \voxel \in \setVoxel_{\structure}, \ \structure\in \setStructure\\
	\label{e:plpr3} & \ \dose_{\tumor \voxel} + \doseUnder_{\tumor \voxel}\geq \targetDose_{\tumor} && \quad \voxel \in \setVoxel_{\tumor}, \ \tumor\in \setTumor\\
	\label{eq:sdior3} & \ \tcolK{\botOpt_{\isocenter} \geq \sum_{\sectorSize \in \setSectorSize} \beamWeight_{\isocenter \sector \sectorSize}}, && \quad \tcolK{\isocenter \in \setIsocenter, \ \sector \in \setSector}\\
	\label{e:sdior6} & \ \tcolK{\sum_{\isocenter \in \setIsocenter} \isoVar_{\isocenter} \leq \UBNumIsocenter} \\
	\label{e:sdior7} & \ \tcolK{\beamWeight_{\isocenter\sector\sectorSize} \leq \isoVar_{\isocenter}\bigM} && \quad \tcolK{\isocenter \in \setIsocenter, \ \sector \in \setSector, \ \sectorSize \in \setSectorSize} \\
	\label{e:sdior8} & \ \tcolK{\isoVar_{\isocenter} \in \{0,1\}}, \ \botOpt_{\isocenter} \geq 0  && \quad \tcolK{\isocenter \in \setIsocenter}\\
	\label{e:sdior4} & \ \beamWeight_{\isocenter\sector\sectorSize} \geq 0  && \quad \isocenter \in \setIsocenter, \ \sector \in \setSector, \ \sectorSize \in \setSectorSize\\
	\label{e:sdior5} & \ \doseUnder_{\structure \voxel} \geq 0, \ \ \doseOver_{\structure \voxel} \geq 0 && \quad \voxel \in \setVoxel_{\structure}, \ \structure\in \setStructure
	\end{align}
	\end{subequations}
\end{small}

\noindent The objective function and the constraint sets of $\modelLPone$ and $\modelSDIOone$ are identical except for the constraints in $\modelSDIOone$ that govern isocenter selection and irradiation times from the selected isocenters. Specifically, constraints \eqref{e:sdior6} ensure that at most $\UBNumIsocenter$ isocenters are selected, and constraints \eqref{e:sdior7} are big-$M$ type constraints guaranteeing that dose can only be delivered through selected isocenters. In order to better represent the total treatment time, we also incorporate the number of isocenters used in the treatment plan to the objective function \eqref{e:sdior0}. Specifically, we take the time spent switching between two isocenters as $\isoCoef$, which makes the total time spent for the isocenters to be $\isoCoef \sum_{\isocenter \in \setIsocenter} \isoVar_{\isocenter}$.

\subsection{Benders Decomposition for SDIO}
Benders decomposition is a commonly used technique for solving certain large-scale optimization problems \citep{Benders1962}. Instead of considering all decision variables and constraints of an optimization problem simultaneously, Benders decomposition partitions the problem into smaller problems. In particular, Benders decomposition algorithm iteratively solves a master problem, which assigns tentative values for the master problem variables, and potentially multiple subproblems that are obtained by fixing the master problem variables to these tentative values. At each iteration, the subproblem solutions provide information on the value of master variables, which are expressed as Benders cuts for the master problem. For the SDIO problem, we formulate the master problem in the Benders decomposition as follows:\\

\vspace*{-1.6cm}
\begin{small}
	\begin{subequations}
	\begin{align}
	\label{e:mpr0} \hhsp{-0.1cm} \text{\modelSDIOMPone}: \min & \ \sum_{\oar\in \setOAR} \sum_{\voxel \in \setVoxel_{\oar}} \objWeightLP_{\oar}\dose_{\oar \voxel} + \tcolK{\coefBOT \sum_{\isocenter \in \setIsocenter} \big(\botOpt_{\isocenter} + \isoCoef \isoVar_{\isocenter} \big)} + \tcolK{\sum_{\oar \in \setOAR} \masterVar_{\oar}^{o} + \sum_{\tumor \in \setTumor} \masterVar_{\tumor}^{o} + \sum_{\tumor \in \setTumor} \masterVar_{\tumor}^{u}} \hhsp{-4.8cm}\\
	\label{e:mpr1} \sthat & \ \dose_{\structure \voxel} = \sum_{\isocenter \in \setIsocenter}\sum_{\sector \in \setSector}\sum_{\sectorSize \in \setSectorSize}\doseMat_{\structure\voxel\isocenter\sector\sectorSize} \beamWeight_{\isocenter\sector\sectorSize} && \quad \voxel \in \setVoxel_{\structure}, \ \structure \in \setStructure\\
		\label{e:mpr2} & \ \tcolK{\botOpt_{\isocenter} \geq \sum_{\sectorSize \in \setSectorSize} \beamWeight_{\isocenter \sector \sectorSize}}, && \quad \tcolK{\isocenter \in \setIsocenter, \ \sector \in \setSector}\\
	\label{e:mpr3} & \ \tcolK{\sum_{\isocenter \in \setIsocenter} \isoVar_{\isocenter} \leq \UBNumIsocenter} \\
	\label{e:mpt4} & \ \tcolK{\beamWeight_{\isocenter\sector\sectorSize} \leq \isoVar_{\isocenter}\bigM} && \quad \tcolK{\isocenter \in \setIsocenter, \ \sector \in \setSector, \ \sectorSize \in \setSectorSize} \\
	\label{e:mpt5} & \ \beamWeight_{\isocenter\sector\sectorSize} \geq 0  && \quad \isocenter \in \setIsocenter, \ \sector \in \setSector, \ \sectorSize \in \setSectorSize\\
 \label{e:mpt6} & \ \tcolK{\isoVar_{\isocenter} \in \{0,1\}, \ \botOpt_{\isocenter} \geq 0} && \quad \tcolK{\isocenter \in \setIsocenter}
\end{align}
\end{subequations}
\end{small}

\noindent where $\masterVar_{\oar}^{o}$, $\masterVar_{\tumor}^{o}$, and $\masterVar_{\tumor}^{u}$ represent, respectively, the optimal objective values of the subproblems $SP_{\oar}^o(\hat{\beamWeight})$, $SP_{\tumor}^o(\hat{\beamWeight})$, and $SP_{\tumor}^u(\hat{\beamWeight})$, which can be formulated as\\

\vspace*{-1.6cm}
\begin{small}
	\begin{subequations}
	\begin{align*}
	\hhsp{-0.1cm} \text{$SP_{\oar}^o(\hat{\beamWeight})$}: \min & \ \sum_{\voxel \in \setVoxel_{\oar}} \objWeightLP_{\oar}\doseOver_{\oar \voxel} \hhsp{-4.8cm}\\
	\sthat & \ \sum_{\isocenter \in \setIsocenter}\sum_{\sector \in \setSector}\sum_{\sectorSize \in \setSectorSize}\doseMat_{\oar\voxel\isocenter\sector\sectorSize} \hat{\beamWeight}_{\isocenter\sector\sectorSize} - \targetDose_{\oar}^{\max} \leq \doseOver_{\oar \voxel} && \quad \voxel \in \setVoxel_{\oar}\\
	& \ \doseOver_{\oar \voxel} \geq 0 && \quad \voxel \in \setVoxel_{\oar}
	\end{align*}
\end{subequations}
\end{small}

\vspace*{-1.6cm}
\begin{small}
	\begin{subequations}
	\begin{align*} 
	\hhsp{-0.1cm} \text{$SP_{\tumor}^o(\hat{\beamWeight})$}: \min & \ \sum_{\voxel \in \setVoxel_{\tumor}} \objWeightLPTumOver_{\tumor}\doseOver_{\tumor \voxel} \hhsp{-4.8cm}\\
	\sthat & \ \sum_{\isocenter \in \setIsocenter}\sum_{\sector \in \setSector}\sum_{\sectorSize \in \setSectorSize}\doseMat_{\tumor\voxel\isocenter\sector\sectorSize} \hat{\beamWeight}_{\isocenter\sector\sectorSize} - \targetDose_{\tumor}^{\max} \leq \doseOver_{\tumor \voxel} && \quad \voxel \in \setVoxel_{\tumor}\\
	& \ \doseOver_{\tumor \voxel} \geq 0 && \quad \voxel \in \setVoxel_{\tumor}
	\end{align*}
\end{subequations}
\end{small}

\vspace*{-1.6cm}
\begin{small}
	\begin{subequations}
	\begin{align*} 
	\hhsp{-0.1cm} \text{$SP_{\tumor}^u(\hat{\beamWeight})$}: \min & \ \sum_{\voxel \in \setVoxel_{\tumor}} \objWeightLPTumUnder_{\tumor}\doseUnder_{\tumor \voxel} \hhsp{-4.8cm}\\
	\sthat & \  -\sum_{\isocenter \in \setIsocenter}\sum_{\sector \in \setSector}\sum_{\sectorSize \in \setSectorSize}\doseMat_{\tumor\voxel\isocenter\sector\sectorSize} \hat{\beamWeight}_{\isocenter\sector\sectorSize} + \targetDose_{\tumor} \leq \doseUnder_{\tumor \voxel} && \quad \voxel \in \setVoxel_{\tumor}\\
	& \ \doseUnder_{\tumor \voxel} \geq 0 && \quad \voxel \in \setVoxel_{\tumor}
	\end{align*}
\end{subequations}
\end{small}

\noindent We can solve $SP_{\oar}^o(\hat{\beamWeight}), \ SP_{\tumor}^o(\hat{\beamWeight}), \ \text{and}\ SP_{\tumor}^u(\hat{\beamWeight})$ by using their duals, which are formulated as follows:\\

\vspace*{-1.6cm}
\begin{small}
	\begin{subequations}
	\begin{align*}
	\hhsp{-0.1cm} \text{$DSP_{\oar}^o(\hat{\beamWeight})$}: \max & \ \sum_{\voxel \in \setVoxel_{\oar}} \dualMaxVar_{\oar\voxel}\big( \sum_{\isocenter \in \setIsocenter}\sum_{\sector \in \setSector}\sum_{\sectorSize \in \setSectorSize}\doseMat_{\oar\voxel\isocenter\sector\sectorSize} \hat{\beamWeight}_{\isocenter\sector\sectorSize} - \targetDose_{\oar}^{\max}\big) \hhsp{-1.8cm}\\
\sthat & \ 0 \leq \dualMaxVar_{\oar\voxel} \leq \objWeightLP_{\oar} && \quad \voxel \in \setVoxel_{\oar}
\end{align*}
\end{subequations}
\end{small}

\vspace*{-1.6cm}
\begin{small}
	\begin{subequations}
	\begin{align*}
	\hhsp{-0.1cm} \text{$DSP_{\tumor}^o(\hat{\beamWeight})$}: \max & \ \sum_{\voxel \in \setVoxel_{\tumor}} \dualMaxVar_{\tumor\voxel}\big( \sum_{\isocenter \in \setIsocenter}\sum_{\sector \in \setSector}\sum_{\sectorSize \in \setSectorSize}\doseMat_{\tumor\voxel\isocenter\sector\sectorSize} \hat{\beamWeight}_{\isocenter\sector\sectorSize} - \targetDose_{\tumor}^{\max}\big) \hhsp{-1.8cm}\\
	\sthat & \ 0 \leq \dualMaxVar_{\tumor\voxel} \leq \objWeightLPTumOver_{\tumor} && \quad \voxel \in \setVoxel_{\tumor}
	\end{align*}
\end{subequations}
\end{small}

\vspace*{-1.6cm}
\begin{small}
	\begin{subequations}
	\begin{align*}
	\hhsp{-0.1cm} \text{$DSP_{\tumor}^u(\hat{\beamWeight})$}: \max & \ \sum_{\voxel \in \setVoxel_{\tumor}} \dualMinVar_{\tumor\voxel}\big( \targetDose_{\tumor} -  \sum_{\isocenter \in \setIsocenter}\sum_{\sector \in \setSector}\sum_{\sectorSize \in \setSectorSize}\doseMat_{\tumor\voxel\isocenter\sector\sectorSize} \hat{\beamWeight}_{\isocenter\sector\sectorSize}\big) \hhsp{-1.8cm}\\
\sthat & \ 0 \leq \dualMinVar_{\tumor\voxel} \leq \objWeightLPTumUnder_{\tumor} && \quad \voxel \in \setVoxel_{\tumor}
\end{align*}
\end{subequations}
\end{small}

It is important to note that these dual subproblems can be solved via inspection, which is beneficial for the performance of the Benders decomposition. For instance, for $DSP_{\oar}^o(\hat{\beamWeight})$, we can find the values for $\dualMaxVar_{\oar\voxel}$ variables by using the following routine:
\begin{itemize}
	\item[] \hhspe if \ \ $\disps \sum_{\isocenter \in \setIsocenter}\sum_{\sector \in \setSector}\sum_{\sectorSize \in \setSectorSize}\doseMat_{\oar\voxel\isocenter\sector\sectorSize} \hat{\beamWeight}_{\isocenter\sector\sectorSize} \geq \targetDose_{\oar}^{\max}$:
	\item[] \hhspe \hhspe set $\dualMaxVar_{\oar\voxel} = \objWeightLP_{\oar}$
	
	\vvspnt
	\item[] \hhspe else:
	\vvspnt
	\item[] \hhspe \hhspe set $\dualMaxVar_{\oar\voxel} = 0$
\end{itemize}

\noindent The same solution approach is also valid for $DSP_{\tumor}^o(\hat{\beamWeight})$ and $DSP_{\tumor}^u(\hat{\beamWeight})$. After getting the optimal solutions of the dual subproblems, Benders decomposition iteratively adds the following cuts to the master problem in the case that they are violated:

\vspace*{-1.6cm}
	\begin{subequations}
	\begin{align}
	\label{e:bd2cut0} \tcolK{\masterVar_{\oar}^{o}} & \ \tcolK{\geq \sum_{\voxel \in \setVoxel_{\oar}} \hat{\dualMaxVar}_{\oar\voxel}\big( \sum_{\isocenter \in \setIsocenter}\sum_{\sector \in \setSector}\sum_{\sectorSize \in \setSectorSize}\doseMat_{\oar\voxel\isocenter\sector\sectorSize} \beamWeight_{\isocenter\sector\sectorSize} - \targetDose_{\oar}^{\max}\big)} && \quad \oar \in \setOAR\\
	\label{e:bd2cut1} \tcolK{\masterVar_{\tumor}^{o}} & \ \tcolK{\geq \sum_{\voxel \in \setVoxel_{\tumor}} \hat{\dualMaxVar}_{\tumor\voxel}\big( \sum_{\isocenter \in \setIsocenter}\sum_{\sector \in \setSector}\sum_{\sectorSize \in \setSectorSize}\doseMat_{\tumor\voxel\isocenter\sector\sectorSize} \beamWeight_{\isocenter\sector\sectorSize} - \targetDose_{\tumor}^{\max}\big)} && \quad \tumor \in \setTumor\\
		\label{e:bd2cut2} \tcolK{\masterVar_{\tumor}^{u}} & \ \tcolK{\geq \sum_{\voxel \in \setVoxel_{\tumor}} \hat{\dualMinVar}_{\tumor\voxel}\big( \targetDose_{\tumor} - \sum_{\isocenter \in \setIsocenter}\sum_{\sector \in \setSector}\sum_{\sectorSize \in \setSectorSize}\doseMat_{\tumor\voxel\isocenter\sector\sectorSize} \beamWeight_{\isocenter\sector\sectorSize}\big)} && \quad \tumor \in \setTumor
	\end{align}
\end{subequations}

\noindent The decomposition algorithm stops when no more violated cuts can be found. Note that there are alternative ways to decompose $\modelSDIOone$ into a master problem and subproblems. For instance, we can have a master problem only with $\isoVar$-variables, and a subproblem consisting of the remaining variables and constraints of $\modelSDIOone$. However, our preliminary analyses indicate that such a decomposition performs worse than the proposed approach.

\subsection{Generating initial feasible solutions for SDIO}\label{sec:InitFeas}
The availability of a high-quality initial feasible solution can help improve the solvability of $\modelSDIOone$ since it provides a good upper bound, which allows the solvers to fathom more nodes by bound and apply strategies such as reduced cost fixing. We use the LP relaxation of the SDIO model to generate an initial feasible solution as follows: 
\begin{itemize}
	\item[1.] Solve the LP relaxation of SDIO model \eqref{e:sdior}, and store the optimal solution of $\isoVar$-variables to the vector $\acute{\isoVar}$.
	\item[2.] Designate $\UBNumIsocenter$ isocenter indices with the largest values from $\acute{\isoVar}$ as the selected isocenters $\acute{\setIsocenter}$. Create a binary vector, $\hat{\isoVar}$, of size $|\setIsocenter|$, and set the components of $\hat{\isoVar}$ corresponding to selected isocenters ($\acute{\setIsocenter}$) to one and the rest of the components to zero.
	\item[3.] Solve SDIO model \eqref{e:sdior} with $\isoVar$-variables fixed to corresponding values in $\hat{\isoVar}$, and store the optimal values for $\beamWeight, \ \doseOver, \ \doseUnder, \ \text{and} \ \botOpt$ variables to $\hat{\beamWeight}, \ \hat{\doseOver}, \ \hat{\doseUnder}, \ \text{and} \ \hat{\botOpt}$ vectors, respectively.
	\item[4.] Solve $DSP_{\oar}^o(\hat{\beamWeight}),\ DSP_{\tumor}^o(\hat{\beamWeight}),\ \text{and}\ DSP_{\tumor}^u(\hat{\beamWeight})$, and store the optimal objective values for these subproblems as $\hat{\masterVar}_{\oar}^{o}, \ \hat{\masterVar}_{\tumor}^{o}, \ \hat{\masterVar}_{\tumor}^{u}$.
\end{itemize}

\noindent We obtain an initial solution to $\modelSDIOMPone$ as the solution set $(\hat{\isoVar}, \hat{\beamWeight}, \hat{\botOpt}, \hat{\masterVar}_{\oar}^{o}, \ \hat{\masterVar}_{\tumor}^{o}, \ \hat{\masterVar}_{\tumor}^{u})$. Similarly, an initial solution for $\modelSDIOone$ is obtained as the solution set $(\hat{\isoVar}, \hat{\beamWeight}, \hat{\botOpt}, \ \hat{\doseOver}, \ \hat{\doseUnder})$. Note that as long as we construct a vector $\hat{\isoVar}$ with at most $\UBNumIsocenter$ ones and $|\setIsocenter| - \UBNumIsocenter$ zeros in Step 2, we always end up with a feasible solution for $\modelSDIOone$ and $\modelSDIOMPone$ following steps 3 and 4. As such, randomly picking $\UBNumIsocenter$ isocenters in Step 2 will lead to a feasible solution as well. We observe that these randomly generated feasible solutions usually do not lead to tight upper bounds for $\modelSDIOone$ and $\modelSDIOMPone$. On the other hand, LP-based heuristics that rely on the LP relaxations of mixed-integer programming models have been successfully used for several problems in the literature (e.g., see \citet{Danna2005} and \citet{Toledo2015}).

\subsection{Experimental setup}\label{sec:expSetup}
\textit{Implementation details.} We implement our algorithms using ILOG CPLEX 12.7 running on a PC with a 3.00 GHz Quad Core Intel Xeon CPU and 16 GB RAM. For the Benders algorithm, we observe that repeatedly solving $\modelSDIOMPone$ --- which is a mixed-integer programming problem --- to optimality, adding cuts and re-solving it can be very expensive from a computational point of view. In order to improve the efficiency of the algorithm, we can interrupt the branch-and-bound solution process of $\modelSDIOMPone$ each time the solver finds a solution, generate Benders cuts, and resume solution process (i.e., operate over a single branch-and-bound tree). In our computational tests, this approach outperforms solving $\modelSDIOMPone$ to optimality in each iteration, adding cuts, and re-optimizing it. A similar approach is used in several studies including \citet{Codato2006} and \citet{Taskin2010a}. We use CPLEX's callback functions to solve our model using a single branch-and-bound tree.

Specifications of the test cases are provided in Table \ref{tb:data}. We observe that the objective function weights ($\objWeightLP$) may significantly affect the solvability of SDIO as well as quality of the resulting plans. Accordingly, we use a simulated annealing algorithm to select the $\objWeightLP$ values in our models (see \citet{Cevik2018} for more details). In addition, we take $\bigM$ as 50 in our numerical experiments. We provide a sensitivity analysis on $\objWeightLP$ and $\bigM$ parameters in Section \ref{sec:ressSA}. We refer to solving the SDIO model \eqref{e:sdior} using CPLEX as ``Cplex'', and using Benders decomposition as ``Benders''. For both Cplex and Benders, we generate initial feasible solutions using the procedure described in Section \ref{sec:InitFeas}. We also consider a quick solution generation method (UBLB), which obtains a lower bound on the optimal value using the LP relaxation of SDIO \eqref{e:sdior} and an upper bound using the generated initial feasible solutions.

\setlength{\tabcolsep}{3pt}
\begin{table}[h]
\centering
\caption{Specifications of the test cases}
\begin{small}
\scalebox{0.9}{
\newcommand{\mcp}{\multicolumn{1}{p{1.95cm}}}
\newcommand{\mcps}{\multicolumn{1}{p{1.3cm}}}
\newcommand{\mro}{\multirow{2}{0.9cm}}
\newcommand{\mrt}{\multirow{2}{0.7cm}}
\newcommand{\cnt}{\centering}
\newcommand{\rrt}{\raggedright}
\begin{threeparttable}
\begin{tabular}{lcccccccc}
\toprule
\toprule
\mro{\rrt Case}  & \mcps{\cnt Tumor\\ Rx (Gy)} & \mcp{\cnt OAR dose\\ limit (Gy)} & \mcp{\cnt Number of\\ isocenters}& \mcp{\cnt Tumor \\ volume (cm$^3$)} & \mcp{\cnt Number of\\ tumor voxels} & \mcp{\cnt Number of\\ ring voxels} & \mcp{\cnt Number of\\ OAR voxels} & \mcp{\cnt Total number \\ of voxels} \\
\toprule
1 & 14 & \{15, 15\} & 24 & 4.52 & 4,524 & 695 & 10,068 & 15,287 \\
2 & 24 & \{15, 8\} & 25 & 5.23 & 5,234 & 6,657 & 8,934 & 20,825 \\
3 & 12.5 & \{15, 15\} & 19 & 1.35 & 1,347 & 3,089 & 5,844 & 10,280 \\
4 & 17 & - & 34 & 12.83 & 12,829 & 18,172 & - & 31,001 \\
5 & 20 & - & 8 & 1.63 & 1,628 & 3,809 & - & 5,437 \\
6 & 12 & \{15, 11.5\} & 20 & 2.6 & 2,601 & 4,077 & 8,108 & 14,786 \\
7 & 12.5 & \{15, 7\} & 33 & 2.52 & 2,521 & 4,235 & 7,643 & 14,399 \\
8 & 18 & - & 16 & 3.45 & 3,447 & 6,833 & - & 10,280 \\
\bottomrule
\bottomrule
\end{tabular}
\begin{tablenotes}
\item[]	Rx: Prescribed dose
\end{tablenotes}
\end{threeparttable}

}
\end{small}
\label{tb:data}
\end{table}

In the clinical plans, isocenter locations are usually manually selected to span the tumor volume. Ideally any coordinate within the tumor volume can be a candidate for an isocenter location. However, even when we only take the tumor voxels to be candidate isocenter locations, we may end up with thousands of eligible isocenters, which would lead to computationally intractable SDIO models. As such, also to promote the automation in treatment plan generation, we use grassfire and sphere packing algorithms (GSP) to determine the eligible set of isocenters ($\setIsocenter$). If GSP generates a low number of isocenters, we use random sampling to increase the number of isocenters to 50 or 100 (depending on the experiment and instance).

\section{Results}
\subsection{Performance of the proposed approach}
\begin{figure*}[h]
	\hhsp{-0.4em}
	\begin{adjustbox}{minipage=\linewidth,scale=1.00}
	\subfloat[\label{fig:rperfgap1}]{\includegraphics[width=0.5\textwidth]{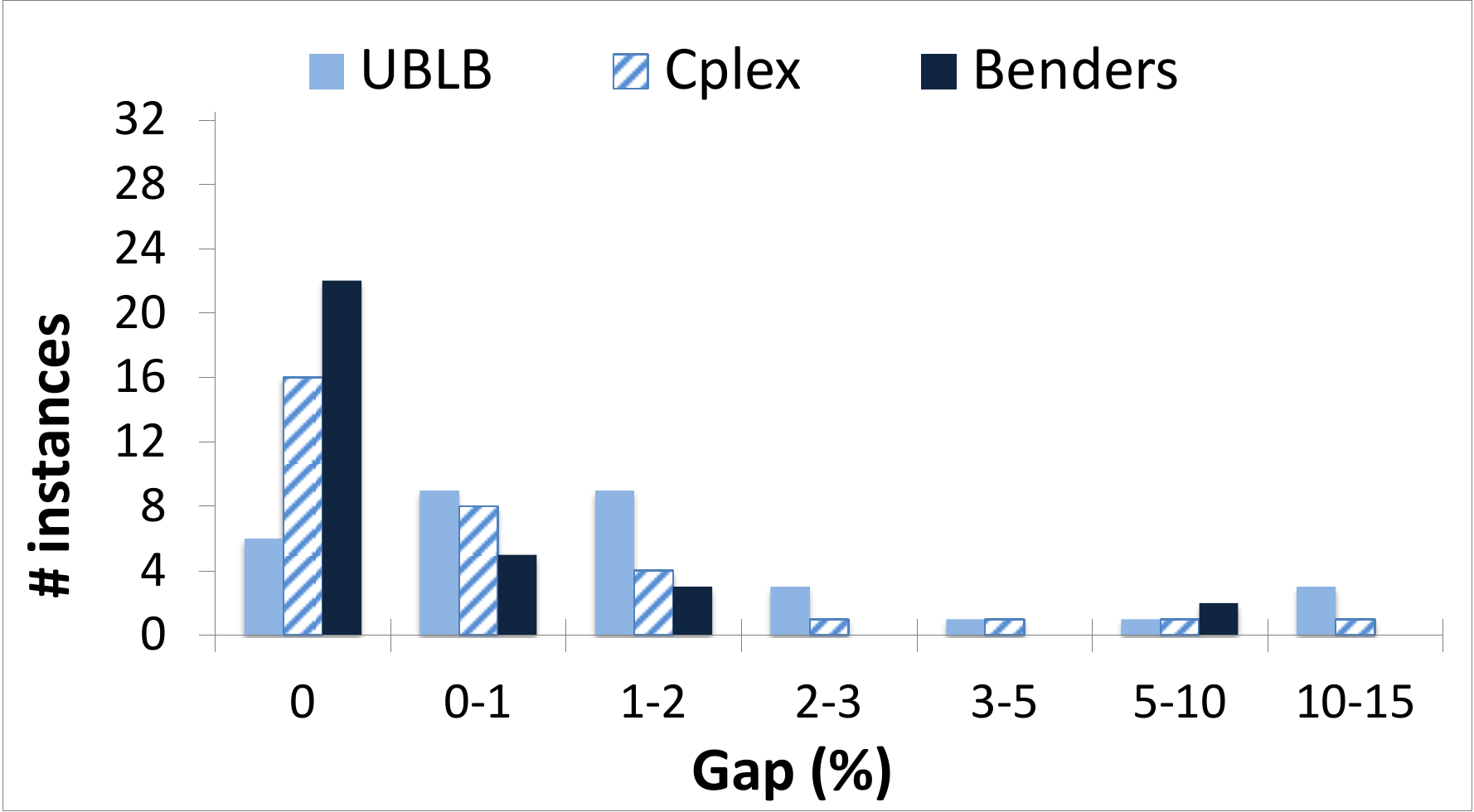}} 
	\quad
	\subfloat[\label{fig:rperftime1}]{\includegraphics[width=0.5\textwidth]{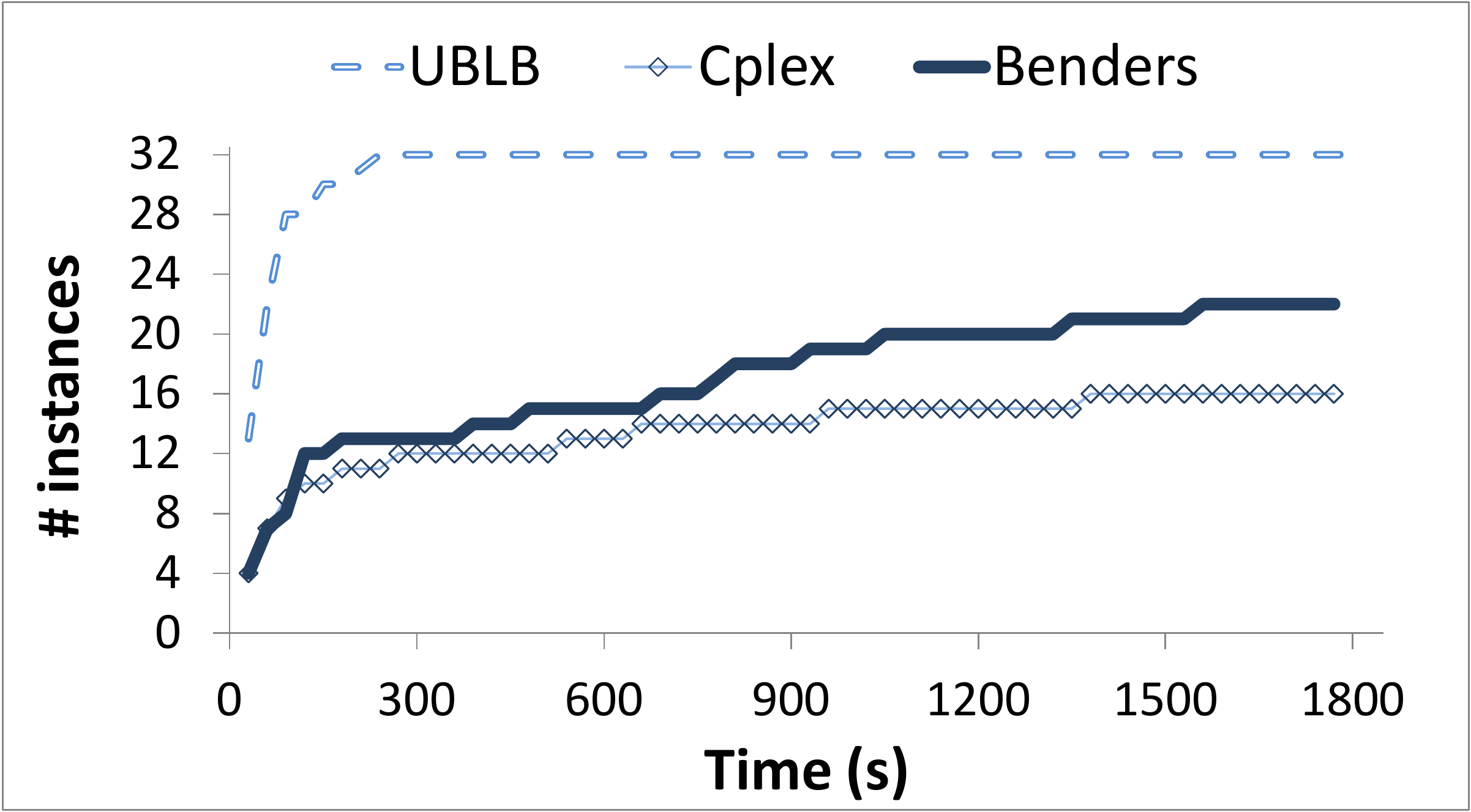}}
	\end{adjustbox} 
	\caption{Performance analysis results.}
	\label{fig:rperf1}
\end{figure*}

We compare performances of Cplex, Benders and UBLB methods in terms of convergence to optimal solutions (see Figure \ref{fig:rperf1}). We consider all eight cases using varying sizes of isocenter sets ($\setIsocenter$) and limits on the number of isocenters ($\UBNumIsocenter$), which leads to 32 instances. In Figure \ref{fig:rperfgap1}, we present the frequency histograms of the optimality gaps (i.e., the percentage difference between best upper and lower bounds) obtained by each method. Figure \ref{fig:rperftime1} is a cumulative distribution plot showing the number of instances solved within a given time limit, where the maximum allowed time limit is 1800 seconds. Note that while UBLB data shown in Figure \ref{fig:rperftime1} represent the time for algorithm to converge (not necessarily to an optimal solution), Cplex and Benders data represent the time to converge to optimal solutions.

We find that the UBLB method frequently generates high quality solutions within a reasonable amount of time. However, UBLB may lead to large optimality gaps for some instances (e.g., in four of the instances, the optimality gap is larger than 10\%). The results reported in Figure \ref{fig:rperf1} suggest that Benders outperforms Cplex both in terms of solution times and the number of instances solved to optimality. Accordingly, we use Benders decomposition in the remainder of the experiments.

\subsection{Impact of isocenter levels}
We investigate the impact of isocenter counts on treatment plan quality by solving the SDIO model using different isocenter limits ($\UBNumIsocenter$). For each case, we fix the number of isocenters to be 50 as described in Section \ref{sec:expSetup}. We first consider a specific case (Case 6) and observe the change in the plan quality as the number of isocenters used in the treatment plan decreases (see Table \ref{tb:rincr1}). We observe that it is possible to obtain high quality treatment plans using much fewer isocenters than the clinical plans. In particular, $\modelDualLPone$ generates a better plan with a lower max OAR dose, higher Paddick index (PCI), comparable gradient index (GI) and lower beam-on-time (BOT) value compared to Clinical plan by using fewer isocenters (20 vs 17). We note that using fewer isocenters only marginally increases the $\modelSDIOone$ objective value, which represents total penalty associated with the treatment plan. Moreover, compared to the treatment plan obtained by $\modelDualLPone$, using seven isocenters only marginally reduces the PCI (0.84 vs 0.82), and slightly increases the BOT (22.9 vs 23.5).

\begin{table}[h]
\centering
\caption{Impact of isocenter limits on treatment plan quality using Case 6}
\begin{small}
\setlength{\tabcolsep}{8pt}
\scalebox{0.9}{
\newcommand{\mcp}{\multicolumn{1}{p{1.8cm}}}
\newcommand{\mcs}{\multicolumn{1}{p{0.8cm}}}
\newcommand{\cnt}{\centering}
\begin{threeparttable}
\begin{tabular}{rrcccccccccc}
\toprule
Method & $\UBNumIsocenter$ & \mcp{\cnt max OAR\\ dose (Gy)} & PCI & GI & \mcs{\cnt BOT \\ (min)} & \mcs{\cnt num. \\ iso.} & \mcs{\cnt num. \\ shots} & UB & LB & Gap(\%) & CPU(s) \\
\midrule
Clinical & - & \{15.0, 11.5\} & 0.80 & 3.00 & 90.7 & 20 & 20 & - & - & - & - \\
DualLP & - & \{13.6, 6.9\} & 0.84 & 3.09 & 32.0 & 17 & 27 & 184.6 & 184.6 & 0.00 & 13.6 \\
Benders & 14 & \{13.6, 6.9\} & 0.84 & 3.08 & 31.6 & 14 & 22 & 184.7 & 184.6 & 0.01 & 82.2 \\
Benders & 11 & \{13.6, 6.9\} & 0.84 & 3.06 & 31.2 & 11 & 22 & 184.8 & 184.7 & 0.10 & 160.0 \\
Benders & 7 & \{13.8, 6.9\} & 0.82 & 2.99 & 32.9 & 7 & 17 & 186.4 & 186.2 & 0.10 & 1041.5 \\
\bottomrule
\end{tabular}
\end{threeparttable}
}
\end{small}
\label{tb:rincr1}
\end{table}

We then determine two levels of isocenter limits ($\UBNumIsocenter$) for each test case, which are obtained by taking 80\% ($\UBNumIsocenter_1$) and 60\% ($\UBNumIsocenter_2$) of the number of isocenters that are used in the corresponding treatment plans generated by $\modelDualLPone$. Then, we compare the quality of the treatment plans generated by $\modelDualLPone$ and $\modelSDIOone$ with isocenter limits $\UBNumIsocenter_1$ and $\UBNumIsocenter_2$ (see Table \ref{tb:rsumm1}). These results show that the plans generated by the $\modelDualLPone$ and $\modelSDIOone$ are of better quality compared to clinical plans. Moreover, we observe that $\modelSDIOone$ is able to generate similar quality plans to $\modelDualLPone$ by using significantly fewer isocenters. More detailed results are provided in Appendix \ref{sec:app1}.

\thispagestyle{empty}
\begin{landscape}
\setlength{\tabcolsep}{7.5pt}
\renewcommand{\arraystretch}{1.29}
\vspace*{\fill}
\begin{table}[!ht]
\centering
\caption{Impact of isocenter limits on treatment plan quality}
\scalebox{0.9}{
\newcommand{\mcp}{\multicolumn{1}{p{1.8cm}}}
\newcommand{\mcs}{\multicolumn{1}{p{0.8cm}}}
\newcommand{\mro}{\multirow{2}{0.6cm}}
\newcommand{\mrt}{\multirow{2}{0.7cm}}
\newcommand{\cnt}{\centering}
\newcommand{\rrt}{\raggedright}
\begin{threeparttable}
\begin{tabular}{rccccclccccclccccc}
\toprule
\toprule
 & \multicolumn{5}{c}{$\modelDualLPone$} &  & \multicolumn{5}{c}{$\modelSDIOone$-$\UBNumIsocenter_1$} &  & \multicolumn{5}{c}{$\modelSDIOone$-$\UBNumIsocenter_2$} \\
\cmidrule(lr){2-6} \cmidrule(lr){8-12} \cmidrule(lr){14-18}
Case & \mcp{\cnt max OAR \\ dose (Gy)} & PCI & \mcs{\cnt BOT \\ (min)} & \mcs{\cnt num. \\ iso.} & \mcs{\cnt num. \\ shots} &  & \mcp{\cnt max OAR \\ dose (Gy)} & PCI & \mcs{\cnt BOT \\ (min)} & \mcs{\cnt num. \\ iso.} & \mcs{\cnt num. \\ shots} &  & \mcp{\cnt max OAR \\ dose (Gy)} & PCI & \mcs{\cnt BOT \\ (min)} & \mcs{\cnt num. \\ iso.} & \mcs{\cnt num. \\ shots} \\
\midrule
1 & \{\textbf{16.4}, 3.3\} & \textbf{0.94} & \textbf{57.6} & 30 & 50 &  & \{\textbf{16.2}, 3.4\} & \textbf{0.93} & \textbf{56.3} & \textbf{24} & 51 &  & \{\textbf{16.2}, 3.1\} & \textbf{0.93} & \textbf{54.7} & \textbf{18} & 44 \\
2 & \{17.5, \textbf{7.8}\} & \textbf{0.79} & \textbf{100.8} & \textbf{23} & 27 &  & \{17.5, \textbf{7.5}\} & \textbf{0.79} & \textbf{100.2} & \textbf{21} & \textbf{25} &  & \{17.4, \textbf{7.5}\} & \textbf{0.79} & \textbf{100.7} & \textbf{17} & \textbf{21} \\
3 & \{15.2, \textbf{10.7}\} & \textbf{0.75} & \textbf{80.1} & 21 & 39 &  & \{15.2, \textbf{10.5}\} & \textbf{0.75} & \textbf{75.7} & \textbf{17} & 31 &  & \{15.5, \textbf{10.4}\} & \textbf{0.72} & \textbf{75.6} & \textbf{13} & 27 \\
4 & - & \textbf{0.76} & \textbf{54.4} & 35 & 60 &  & - & \textbf{0.76} & \textbf{54.6} & \textbf{28} & 50 &  & - & \textbf{0.73} & \textbf{55.5} & \textbf{21} & 48 \\
5 & - & \textbf{0.84} & \textbf{20.1} & 16 & 24 &  & - & \textbf{0.85} & \textbf{20.1} & 13 & 22 &  & - & \textbf{0.84} & \textbf{20.3} & 10 & 17 \\
6 & \{\textbf{13.6}, \textbf{6.9}\} & \textbf{0.84} & \textbf{32.0} & \textbf{17} & 27 &  & \{\textbf{13.6}, \textbf{6.9}\} & \textbf{0.84} & \textbf{31.6} & \textbf{14} & \textbf{22} &  & \{\textbf{13.6}, \textbf{6.9}\} & \textbf{0.84} & \textbf{31.2} & \textbf{11} & 22 \\
7 & \{13.7, \textbf{1.3}\} & \textbf{0.85} & \textbf{43.6} & \textbf{18} & 42 &  & \{13.8, \textbf{1.3}\} & \textbf{0.85} & \textbf{43.3} & \textbf{15} & 35 &  & \{13.6, \textbf{1.3}\} & \textbf{0.84} & \textbf{43.7} & \textbf{11} & \textbf{33} \\
8 & - & \textbf{0.88} & \textbf{42.7} & 21 & 38 &  & - & \textbf{0.88} & \textbf{43.1} & \textbf{16} & 37 &  & - & \textbf{0.88} & \textbf{42.3} & \textbf{13} & 35 \\
\bottomrule
\bottomrule
\end{tabular}
\begin{tablenotes}
\item[] Bold values indicate improvement over the clinical plan.
\end{tablenotes}
\end{threeparttable}

}
\label{tb:rsumm1}
\end{table}
\vspace*{\fill}
\end{landscape}

\begin{figure*}[h]
	\hhsp{-2.4em}
	\begin{adjustbox}{minipage=\linewidth,scale=1.15}
	\subfloat[Dose-volume histogram]{\includegraphics[width=0.5\textwidth]{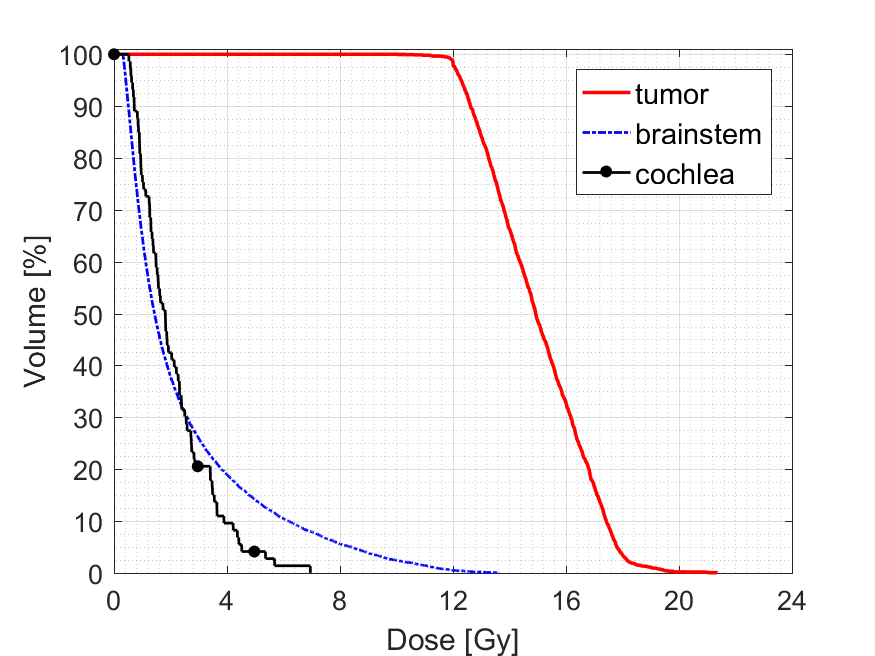}} 
	\hhsp{-1.6em}
	\subfloat[Isodose lines]{\includegraphics[width=0.5\textwidth]{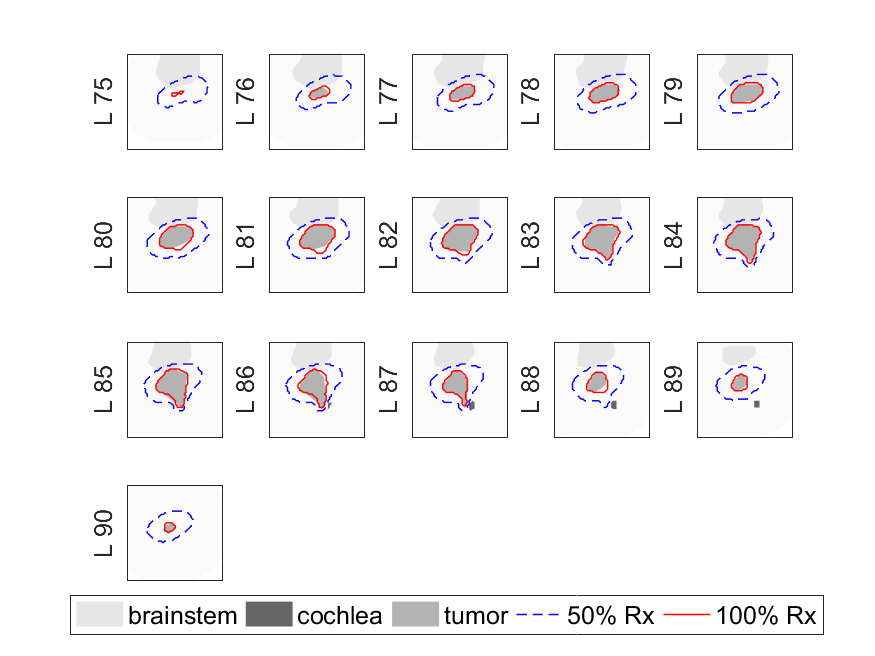}}
	\end{adjustbox}
	\caption{DVH curve and isodose lines for a Case 6 for the treatment plan obtained by $\modelSDIOone$-$\UBNumIsocenter_2$ (11 isocenters).}
	\label{fig:IsodoseAll}
\end{figure*}

We also examine the change in the treatment plan quality when fewer isocenters are used by comparing the DVH curves and the isodose lines. Figure \ref{fig:IsodoseAll} shows that, for a representative case, $\modelSDIOone$-$\UBNumIsocenter_2$ generates highly conformal treatment plans by using 11 isocenters instead of 20 isocenters used in the clinical plan.

\subsection{Sensitivity analysis}\label{sec:ressSA}
We test the impact of the model parameters, namely, \emph{big-M} parameter and \emph{objective function weights}, in the SDIO model on solution and treatment plan qualities using a representative case (Case 6). In Table \ref{tb:rsabigM1}, we consider three isocenters levels (14, 11, and 7) and four $\bigM$ values (20, 50, 100, and 200). We observe that the solvability of the SDIO model is affected by the choice of $\bigM$ value. In particular, we observe that setting $\bigM$ too low (e.g., $\bigM = 20$) may be too restrictive and may lead to a higher total penalty for the generated treatment plans. For instance, for $\UBNumIsocenter = 14$, total penalty (UB) values are 185.7 and 184.7 for $\bigM = 20$ and $\bigM = 50$, respectively. On the other hand, setting $\bigM$ too high (e.g., $\bigM = 100$ or $\bigM = 200$) may lead to increased CPU times as we can see in our experiments with $\UBNumIsocenter = 7$, which can be attributed to potentially worse LP relaxation bounds for the SDIO model. We provide sensitivity analysis results for the impact of objective function weights in Appendix \ref{sec:app2}.

\begin{table}[h]
\centering
\caption{Sensitivity analysis results for big-$M$ value in SDIO using Case 6}
\begin{small}
\setlength{\tabcolsep}{9pt}
\scalebox{0.9}{
\newcommand{\mcp}{\multicolumn{1}{p{1.8cm}}}
\newcommand{\mcs}{\multicolumn{1}{p{0.8cm}}}
\newcommand{\cnt}{\centering}
\begin{threeparttable}
\begin{tabular}{rrcccccccccc}
\toprule
$\UBNumIsocenter$ & $\bigM$ & \mcp{\cnt max OAR \\ dose (Gy)} & PCI & GI & \mcs{\cnt BOT \\ (min)} & \mcs{\cnt num. \\ iso.} & \mcs{\cnt num. \\ shots} & UB & LB & Gap(\%)  & CPU(s) \\
\midrule
14 & 20 & \{13.6, 7.1\} & 0.83 & 3.10 & 30.3 & 14 & 27 & 185.7 & 185.6 & 0.06\% & 178.0 \\
 & 50 & \{13.6, 6.9\} & 0.84 & 3.08 & 31.6 & 14 & 22 & 184.7 & 184.6 & 0.01\% & 81.8 \\
 & 100 & \{13.6, 6.9\} & 0.84 & 3.08 & 31.6 & 14 & 22 & 184.7 & 184.6 & 0.01\% & 82.1 \\
 & 200 & \{13.6, 6.9\} & 0.84 & 3.08 & 31.6 & 14 & 22 & 184.7 & 184.6 & 0.01\% & 83.0 \\
\midrule
11 & 20 & \{13.6, 7.1\} & 0.84 & 3.11 & 30.3 & 11 & 24 & 185.9 & 185.7 & 0.10\% & 291.9 \\
 & 50 & \{13.6, 6.9\} & 0.84 & 3.06 & 31.2 & 11 & 22 & 184.8 & 184.7 & 0.10\% & 159.6 \\
 & 100 & \{13.6, 6.9\} & 0.84 & 3.05 & 31.2 & 11 & 21 & 184.8 & 184.7 & 0.09\% & 144.0 \\
 & 200 & \{13.6, 6.9\} & 0.84 & 3.06 & 31.2 & 11 & 23 & 184.8 & 184.7 & 0.10\% & 140.4 \\
\midrule
7 & 20 & \{14.3, 6.9\} & 0.81 & 2.91 & 32.1 & 7 & 19 & 188.1 & 187.2 & 0.49\% & 1800.0 \\
 & 50 & \{13.8, 6.9\} & 0.82 & 2.99 & 32.9 & 7 & 17 & 186.4 & 186.2 & 0.10\% & 1036.5 \\
 & 100 & \{13.8, 6.9\} & 0.83 & 3.00 & 32.9 & 7 & 17 & 186.4 & 186.2 & 0.10\% & 1533.3 \\
 & 200 & \{13.8, 6.9\} & 0.83 & 3.00 & 32.9 & 7 & 18 & 186.4 & 186.2 & 0.10\% & 1292.0 \\
\bottomrule
\end{tabular}
\end{threeparttable}

}
\end{small}
\label{tb:rsabigM1}
\end{table}

\section{Discussions}
Our SDIO approach to the inverse problem for SRS on Icon{\texttrademark} yields conformal treatment plans that satisfy the clinical objectives. The detailed numerical study indicate that it is possible to achieve high quality treatment plans in terms of PCI and BOT using fewer isocenters. Automatically generated plans by our models are capable of being delivered on the treatment unit as the proposed models are specifically designed to exploit the Icon{\texttrademark}'s automated collimator size changes and couch positioning. Moreover, all possible combinations of collimator sizes in all sectors for a large set of isocenter locations are considered in the optimization models. As such, we present a framework that can be used to guide treatment planners to explore tradeoffs between delivery efficiency in terms of number of isocenters and shots in the plan and dose conformality.

The literature on inverse planning for LGK is limited and only \citet{Ghaffari2017} consider the SDIO problem. In their limited numerical analyses, \citet{Ghaffari2017} show that low BOT values can be achieved while maintaining the quality of the plans by imposing limits on the minimum and maximum irradiation times of the selected isocenters for each sector and sector size. In our study, we directly incorporate the BOT function to the SDIO model, and we use big-$M$ type constraints to ensure that the radiation delivery is only possible through selected isocenters. More importantly, we propose a Benders decomposition scheme to improve the solvability of the SDIO model. Accordingly, we are able to examine the impact of isocenter counts at various levels. On the other hand, we recognize that the efficiency of the solution method for SDIO can be further improved to solve larger instances with higher number of candidate isocenters, which is left for future research. 


It is important to note that, although the clinical objectives can be more easily achieved using a larger number of isocenters, it is preferable to achieve the same goals with fewer isocenters. Although the time spent by LGK to switch between the isocenters is around 6-10 seconds \citep{ElektaWhitePaper2010}, having a large number of isocenters in the treatment plan leads to longer treatment times. In the context of the frameless workflow on the LGK Icon{\texttrademark}, minimizing overall treatment time is of utmost importance since the patient's tolerability to remain still in a mask decreases with time. In the institution that we get our test cases from, the clinicians treat 5-6 frameless cases per day and they make every attempt to keep overall treatment time below 45 minutes since this duration is observed to be the time that patients can tolerate the immobilization device. On the Icon{\texttrademark} system, a cone-beam CT (CBCT) scanner is used for setup verification and subsequently an optical tracking device monitors the patient for motion in real-time. If the patient moves out of tolerance the treatment is immediately paused and the patient must undergo a repeat CBCT and subsequent plan evaluation, a process that adds an additional five minutes to the treatment time \citep{Chung2018}. Since the tolerability of the patient to the immobilization mask decreases with time, it is paramount to the success of treatment that the overall time is minimized. Even a reduction of a few minutes in treatment time may suffice to mitigate the need for additional CBCT scans.


Among the study limitations is the small sample size used for testing the algorithms (eight patient cases). While the clinical cases were selected to represent the challenging cases in terms of shape complexity and/or OAR proximity, more generalized conclusions as to model stratified by tumor type cannot be made with this limited sample size. Accordingly, we aim to further expand our analysis using a larger data set to test the robustness of approach, as well as quantifying the relative gains and benefits of employing our LP, SDIO and UBLB methods.

\section{Conclusions}
We have shown that an integer programming model can be efficiently used to simultaneously optimize isocenter locations and sector duration for the inverse planning with LGK Icon{\texttrademark}. The proposed Benders decomposition scheme is flexible enough to obtain solutions in a clinically viable amount of time even for the cases with a larger number of voxels and candidate isocenter locations. We note that our treatment plans satisfy the clinical objectives while demonstrating high conformity and low BOT. In this regard, our study contributes to automated treatment plan generation. In particular, our approach allows testing the feasibility of using fewer number of isocenters in treatment plans to improve plan efficiency and reduce the total treatment time.

\clearpage
\newpage
\bibliographystyle{apalike}
\bibliography{GK_SDO}

\clearpage
\newpage
\begin{appendices}
\section{\large Appendix}
\vspace{-0.4cm}

\subsection{Detailed numerical results on impact of isocenter limits} \label{sec:app1}

\begin{table}[h]
\centering
\caption{Clinical results}
\begin{small}
\setlength{\tabcolsep}{9pt}
\scalebox{0.9}{
\newcommand{\mcp}{\multicolumn{1}{p{1.8cm}}}
\newcommand{\mcs}{\multicolumn{1}{p{0.8cm}}}
\newcommand{\cnt}{\centering}
\begin{tabular}{rccccccc}
\toprule
Case  & Coverage   & \mcp{\cnt max OAR \\ dose (Gy)} & PCI & GI & \mcs{\cnt BOT \\ (min)}   & \mcs{\cnt num. \\ iso.} & \mcs{\cnt num. \\ shots}\\
\midrule
1 & 0.977 & \{18.1, 2.4\} & 0.910 & 2.59 & 78.0 & 24 & 24 \\
2 & 0.820 & \{9.7, 7.9\} & 0.560 & 2.86 & 120.4 & 25 & 25 \\
3 & 0.970 & \{13.6, 11.7\} & 0.710 & 3.15 & 86.4 & 19 & 19 \\
4 & 0.996 & - & 0.680 & 2.82 & 109.3 & 34 & 34 \\
5 & 0.992 & - & 0.750 & 3.13 & 39.5 & 8 & 8 \\
6 & 0.990 & \{14.6, 11.2\} & 0.800 & 3.00 & 90.7 & 20 & 20 \\
7 & 0.991 & \{12.3, 6.2\} & 0.830 & 2.75 & 90.2 & 33 & 33 \\
8 & 0.989 & - & 0.810 & 3.07 & 65.7 & 16 & 16 \\
\bottomrule
\end{tabular}

}
\end{small}
\label{tb:rappSumm0}
\end{table}

\begin{table}[h]
\centering
\caption{DualLP results}
\begin{small}
\setlength{\tabcolsep}{6pt}
\scalebox{0.9}{
\newcommand{\mcp}{\multicolumn{1}{p{1.8cm}}}
\newcommand{\mcs}{\multicolumn{1}{p{0.8cm}}}
\newcommand{\cnt}{\centering}
\begin{tabular}{rccccccccccc}
\toprule
Case  & Coverage   & \mcp{\cnt max OAR \\ dose (Gy)}     & PCI   & GI   & \mcs{\cnt BOT \\ (min)}   & \mcs{\cnt num. \\ iso.} & \mcs{\cnt num. \\ shots} & UB     & LB     & Gap(\%) & CPU(s)  \\
\midrule
1 & 0.988 & \{16.4, 3.3\} & 0.94 & 2.66 & 57.6 & 30 & 50 & 1870.9 & 1870.9 & 0.00 & 22.5 \\
2 & 0.816 & \{17.5, 7.8\} & 0.79 & 2.89 & 100.8 & 23 & 27 & 3054.8 & 3054.8 & 0.00 & 15.3 \\
3 & 0.968 & \{15.2, 10.7\} & 0.75 & 3.15 & 80.1 & 21 & 39 & 1787.8 & 1787.8 & 0.00 & 8.2 \\
4 & 0.995 & - & 0.76 & 3.28 & 54.4 & 35 & 60 & 382.8 & 382.8 & 0.00 & 44.7 \\
5 & 0.985 & - & 0.84 & 3.36 & 20.1 & 16 & 24 & 20.7 & 20.7 & 0.00 & 7.2 \\
6 & 0.985 & \{13.6, 6.9\} & 0.84 & 3.09 & 32.0 & 17 & 27 & 184.6 & 184.6 & 0.00 & 13.6 \\
7 & 0.987 & \{13.7, 1.3\} & 0.85 & 3.47 & 43.6 & 18 & 42 & 150.5 & 150.5 & 0.00 & 13.6 \\
8 & 0.985 & - & 0.88 & 2.95 & 42.7 & 21 & 38 & 141.4 & 141.4 & 0.00 & 13.5 \\
\bottomrule
\end{tabular}

}
\end{small}
\label{tb:rappSumm1}
\end{table}

\begin{table}[h]
\centering
\caption{SDIO-$\UBNumIsocenter_1$ results}
\begin{small}
\setlength{\tabcolsep}{6pt}
\scalebox{0.9}{
\newcommand{\mcp}{\multicolumn{1}{p{1.8cm}}}
\newcommand{\mcs}{\multicolumn{1}{p{0.8cm}}}
\newcommand{\cnt}{\centering}
\begin{tabular}{rccccccccccc}
\toprule
Case  & Coverage   & \mcp{\cnt max OAR \\ dose (Gy)}     & PCI   & GI   & \mcs{\cnt BOT \\ (min)}   & \mcs{\cnt num. \\ iso.} & \mcs{\cnt num. \\ shots} & UB     & LB     & Gap(\%) & CPU(s)  \\
\midrule
1 & 0.986 & \{16.2, 3.4\} & 0.93 & 2.67 & 56.3 & 24 & 51 & 1871.6 & 1870.8 & 0.04 & 227.8 \\
2 & 0.816 & \{17.5, 7.5\} & 0.79 & 2.89 & 100.2 & 21 & 25 & 3055.0 & 3054.6 & 0.01 & 229.7 \\
3 & 0.968 & \{15.2, 10.5\} & 0.75 & 3.16 & 75.7 & 17 & 31 & 1791.0 & 1789.7 & 0.08 & 87.4 \\
4 & 0.994 & - & 0.76 & 3.28 & 54.6 & 28 & 50 & 383.5 & 382.8 & 0.18 & 1800.0 \\
5 & 0.985 & - & 0.85 & 3.37 & 20.1 & 13 & 22 & 20.7 & 20.7 & 0.01 & 33.9 \\
6 & 0.984 & \{13.6, 6.9\} & 0.84 & 3.08 & 31.6 & 14 & 22 & 184.7 & 184.6 & 0.01 & 82.2 \\
7 & 0.986 & \{13.8, 1.3\} & 0.85 & 3.41 & 43.3 & 15 & 35 & 150.7 & 150.5 & 0.09 & 178.1 \\
8 & 0.984 & - & 0.88 & 2.95 & 43.1 & 16 & 37 & 141.4 & 141.4 & 0.06 & 89.5 \\
\bottomrule
\end{tabular}

}
\end{small}
\label{tb:rappSumm2}
\end{table}

\begin{table}[h]
\centering
\caption{SDIO-$\UBNumIsocenter_2$ results}
\begin{small}
\setlength{\tabcolsep}{6pt}
\scalebox{0.9}{
\newcommand{\mcp}{\multicolumn{1}{p{1.8cm}}}
\newcommand{\mcs}{\multicolumn{1}{p{0.8cm}}}
\newcommand{\cnt}{\centering}
\begin{tabular}{rccccccccccc}
\toprule
Case  & Coverage   & \mcp{\cnt max OAR \\ dose (Gy)}     & PCI   & GI   & \mcs{\cnt BOT \\ (min)}   & \mcs{\cnt num. \\ iso.} & \mcs{\cnt num. \\ shots} & UB     & LB     & Gap(\%) & CPU(s)  \\
\midrule
1 & 0.987 & \{16.2, 3.1\} & 0.93 & 2.64 & 54.7 & 18 & 44 & 1875.8 & 1871.9 & 0.21 & 1800.0 \\
2 & 0.816 & \{17.4, 7.5\} & 0.79 & 2.88 & 100.7 & 17 & 21 & 3056.7 & 3054.6 & 0.07 & 465.5 \\
3 & 0.967 & \{15.5, 10.4\} & 0.72 & 3.07 & 75.6 & 13 & 27 & 1795.9 & 1794.1 & 0.10 & 670.2 \\
4 & 0.995 & - & 0.73 & 3.22 & 55.5 & 21 & 48 & 389.1 & 382.8 & 1.61 & 1800.0 \\
5 & 0.990 & - & 0.84 & 3.25 & 20.3 & 10 & 17 & 20.7 & 20.7 & 0.09 & 38.5 \\
6 & 0.984 & \{13.6, 6.9\} & 0.84 & 3.06 & 31.2 & 11 & 22 & 184.8 & 184.7 & 0.10 & 160.0 \\
7 & 0.987 & \{13.6, 1.3\} & 0.84 & 3.50 & 43.7 & 11 & 33 & 151.4 & 151.3 & 0.10 & 1335.0 \\
8 & 0.984 & - & 0.88 & 2.96 & 42.3 & 13 & 35 & 141.6 & 141.4 & 0.10 & 793.6 \\
\bottomrule
\end{tabular}

}
\end{small}
\label{tb:rappSumm3}
\end{table}

\subsection{Sensitivity analysis results on objective weights} \label{sec:app2}

In Table \ref{tb:rsaobjw1}, we consider four different objective function weight vectors ($\objWeightLP$) and solve the inverse planning problem using DualLP model as well as SDIO model with different isocenter levels. As expected, SDIO solvability and treatment plan quality are affected by the choice of the objective function weight vector. As such, we use a metaheuristic (namely simulated annealing) to choose from the possible $\objWeightLP$-vectors in our numerical experiments. In particular, we take the $\objWeightLP$-vector that best balances the different performance metrics related to generated treatment plans such as PCI, GI and BOT.

\begin{table}[h]
\centering
\caption{Sensitivity analysis results for objective function weights in SDIO using Case 6}
\begin{small}
\setlength{\tabcolsep}{6pt}
\scalebox{0.9}{
\newcommand{\mcp}{\multicolumn{1}{p{1.8cm}}}
\newcommand{\mcs}{\multicolumn{1}{p{0.8cm}}}
\newcommand{\cnt}{\centering}
\begin{threeparttable}
\begin{tabular}{rrrcccccccccc}
\toprule
Method & $\UBNumIsocenter$ & $\objWeightLP$ & \mcp{\cnt max OAR \\ dose (Gy)} & PCI & GI & \mcs{\cnt BOT \\ (min)} & \mcs{\cnt num. \\ iso.} & \mcs{\cnt num. \\ shots} & UB & LB & Gap(\%)  & CPU(s) \\
\midrule
DualLP & - & $\objWeightLP^1$ & \{13.6, 6.9\} & 0.84 & 3.09 & 32.0 & 17 & 27 & 184.6 & 184.6 & 0.00\% & 12.8 \\
 & & $\objWeightLP^2$ & \{13.9, 7.4\} & 0.86 & 3.02 & 37.9 & 22 & 34 & 279.1 & 279.1 & 0.00\% & 14.1 \\
 & & $\objWeightLP^3$ & \{14.2, 6.2\} & 0.86 & 2.97 & 57.9 & 24 & 42 & 317.8 & 317.8 & 0.00\% & 11.2 \\
 & & $\objWeightLP^4$ & \{13.5, 7.5\} & 0.84 & 3.07 & 33.8 & 20 & 31 & 372.6 & 372.6 & 0.00\% & 14.5 \\
\midrule
Benders & 14 & $\objWeightLP^1$ & \{13.6, 6.9\} & 0.84 & 3.08 & 31.6 & 14 & 22 & 184.7 & 184.6 & 0.01\% & 81.7 \\
 &  & $\objWeightLP^2$ & \{14.0, 7.4\} & 0.84 & 2.98 & 36.9 & 14 & 30 & 279.6 & 279.3 & 0.10\% & 731.5 \\
 &  & $\objWeightLP^3$ & \{13.9, 6.4\} & 0.84 & 2.95 & 53.2 & 14 & 40 & 318.9 & 318.6 & 0.10\% & 245.0 \\
 &  & $\objWeightLP^4$ & \{13.5, 7.4\} & 0.84 & 3.05 & 33.5 & 14 & 27 & 372.7 & 372.6 & 0.04\% & 84.5 \\
\midrule
Benders & 11 & $\objWeightLP^1$ & \{13.6, 6.9\} & 0.84 & 3.06 & 31.2 & 11 & 22 & 184.8 & 184.7 & 0.10\% & 160.0 \\
 &  & $\objWeightLP^2$ & \{14.3, 7.5\} & 0.84 & 2.94 & 36.6 & 11 & 28 & 280.0 & 279.7 & 0.10\% & 1396.6 \\
 &  & $\objWeightLP^3$ & \{13.6, 6.3\} & 0.84 & 2.94 & 51.9 & 11 & 41 & 320.3 & 319.8 & 0.14\% & 1800.0\\
 &  & $\objWeightLP^4$ & \{13.4, 7.4\} & 0.84 & 3.03 & 32.7 & 11 & 20 & 373.1 & 372.7 & 0.10\% & 288.4 \\
\midrule
Benders & 7 & $\objWeightLP^1$ & \{13.8, 6.9\} & 0.82 & 2.99 & 32.9 & 7 & 17 & 186.4 & 186.2 & 0.10\% & 1041.4 \\
 &  & $\objWeightLP^2$ & \{13.9, 7.6\} & 0.84 & 2.90 & 34.5 & 7 & 23 & 283.7 & 280.8 & 1.03\% & 1800.0\\
 &  & $\objWeightLP^3$ & \{13.7, 6.2\} & 0.80 & 2.87 & 51.3 & 7 & 26 & 327.3 & 321.5 & 1.76\% & 1800.0\\
 &  & $\objWeightLP^4$ & \{13.9, 7.5\} & 0.82 & 2.97 & 34.2 & 7 & 16 & 376.8 & 375.4 & 0.36\% & 1800.0\\
\bottomrule
\end{tabular}
\begin{tablenotes}
\item[] $\objWeightLP^1$: $\objWeightLP_{\oar} = 10$, $\objWeightLPTumUnder_{\tumor} = 1000$, $\objWeightLPTumOver_{\tumor} = 10$, $\coefBOT = 0.5$
\item[] $\objWeightLP^2$: $\objWeightLP_{\oar} = 20$, $\objWeightLPTumUnder_{\tumor} = 2000$, $\objWeightLPTumOver_{\tumor} = 10$, $\coefBOT = 0.5$
\item[] $\objWeightLP^3$: $\objWeightLP_{\oar} = 25$, $\objWeightLPTumUnder_{\tumor} = 8000$, $\objWeightLPTumOver_{\tumor} = 10$, $\coefBOT = 0.3$
\item[] $\objWeightLP^4$: $\objWeightLP_{\oar} = 20$, $\objWeightLPTumUnder_{\tumor} = 3000$, $\objWeightLPTumOver_{\tumor} = 10$, $\coefBOT = 1.0$
\end{tablenotes}
\end{threeparttable}

}
\end{small}
\label{tb:rsaobjw1}
\end{table}

\end{appendices}

\end {document}